\documentclass[12pt]{iopart}

\pdfoutput=1
\usepackage{iopams}  
\usepackage{graphicx}
\begin{document}
\title[Key issues in High-$T_{c}$]{Key issues in theories of high temperature superconductors}

\author{Sudip Chakravarty}

\address{Department of Physics \& Astronomy, University of California Los Angeles, Los Angeles, California 90095-1547, USA}
\ead{sudip@physics.ucla.edu}
\begin{abstract}
High temperature superconductivity in cuprate superconductors remains an unsolved problem in theoretical physics. The same statement can also be made about a number of other superconductors that have been dubbed unconventional.
What makes these superconductors so elusive is an interesting question in itself. The present manuscript focuses  on the recent magnetic oscillation experiments  and how they fit into the broader picture.   Many aspects of these experiments can be explained by Fermi liquid theory; the key issue is  the extent to which this is true. If true, the entire paradigm developed over the past three decades must be reexamined. A critical analysis of this issue has necessitated a broader analysis of questions about distinct ground states of matter, which may be useful in understanding other unconventional superconductors.

\end{abstract}

\maketitle

\section{Introduction}
There is always a semantic question as to what we mean by {\em key issues}, especially because our perception has evolved over the past  decades. At this instant of time, I will define them to be the set of all issues necessary to elucidate one class of experiments that will be collectively described to be magnetic quantum oscillations~\cite{Doiron-Leyraud:2007,Jaudet:2008,LeBoeuf:2007,Sebastian:2008,Yelland:2008,Sebastian:2009,Audouard:2009,Vignolle:2008,Bangura:2008,Singleton:2009,Rourke:2009,Helm:2009}, which took the community by surprise in 2007.  It is my view that these experiments have the potential to change the landscape of high-$T_{c}$ research, because they hint at a degree of simplicity that was previously unrecognized~\cite{Chakravarty:2008}. Historically, emergent simplicity in a complex field has been a real breakthrough. It is in this spirit that this article is written, and it emphatically does not pretend to be a review. I make no apologies; the reader, if he so chooses, may construe this to be downright malice on my part, but I hope that he will recognize that having a clear focus can be more illuminating than otherwise.

The question posed here is quite simple. Is the normal state from which superconductivity develops a Fermi liquid? Before 2007 such a question would have been at the very least heresy, if not  simply silly. The answer to this question will take us into addressing pseudogap, competing order, non-Fermi liquids and effective Hamiltonians. Why should we be carrying so much baggage with us? The prevailing dogma has been that these superconductors arise from the Mott insulating state of their parent compounds~\cite{Anderson:1997}. Never mind the fact that we do not know what the wave function of a Mott insulator is and  what local Hamiltonian it solves. Never  mind the fact that one can equally well view superconductivity as arising from the overdoped regime of the phase digram, which can not be viewed to be in the proximity of a Mott insulating phase by any stretch of the  imagination~\cite{Kopp:2007}. Nonetheless, the same superconductors can be said to develop from the overdoped regime and are killed as we approach the Mott insulating regime. Paradoxical? You are right to think so. It would be a blast if we could prove that right in the heart of the strange metal-Mott insulating regime, the system in reality is a Fermi liquid---a blow to postmodernism. To confront the postmodernists we need crystal clear experiments that can be interpreted beyond a shadow of doubt. Alas, there lies the rub. The quantum oscillation experiments have not so far risen to that level, although they are quickly achieving that status.

In Section 2,  I outline the importance of recognizing a multiplicity of mechanisms, both competing and cooperating,  that may be responsible for high temperature superconductors (high-$T_{c}$). Section 3 is a brief overview of quantum oscillation measurements. In Section 4, I discuss broken symmetries in the context of high-$T_{c}$ superconductors, and in particular density waves of higher angular momenta. The Section 5 contains the definition of a non-Fermi liquid necessary for an overall understanding. Kohn's theorem regarding quantum oscillations provides a robust perspective in Section 6. A short aside about $\nu=1/2$ quantum Hall effect is provided in Section 7. This brief section serves as an warning that there may be more to it than meets the eye. The concept of Fermi surface reconstruction as an explanation of the quantum oscillation experiments is discussed in Section 8. Section 9 contains some the unresolved puzzles.

\section{The magic bullet}
From the very beginning of high-$T_{c}$ research the presumption has been  that there is a magic bullet that will lead to the secret of these materials. We have witnessed many such attempts: resonating valence bonds, spin fluctuations, interlayer tunneling, gauge theories, stripes, electron-phonon mechanism, quantum criticality,  etc. But the search for a magic bullet  has turned out to be an ineffective strategy and may even be a false premise. This may be unique to the new unconventional superconductors in contrast to  old conventional superconductors. In fact, the conflicts between  experiments, especially in the normal state, and the lack of universality point to the importance of  a set of mechanisms, as opposed to a single mechanism.  Therefore the solution to the high-$T_{c}$ problem may be a combination  of mechanisms. The lack of a unique mechanism may be unsatisfying to a reductionist, but it is not so uncommon from the perspective of emergent phenomena; especially serious bottlenecks exist in a number of outstanding problems in biology or medicine. We basically know what causes superconductivity: a suitable attraction between electrons can form a ground state with a broken global gauge symmetry, with a number of remarkable consequences. What we do not know is how this attraction arises, when it is sizable and when it is not,  and what factors are detrimental to superconductivity. We hope that it is not as complex as problems in biology and an effective low energy Hamiltonian that contains the main ingredients can be found. This principle of multiplicity behooves us to take seriously non-partisan views of
\begin{itemize}
\item Conflicts within theories and experiments. Barring highly idealized many particle systems, it is impossible to find a complete mathematical solution of a complex problem, nor is it necessary. History is replete with such examples. Each approximation must be placed in its proper context and both pro and con should be weighed. Similarly, experiments should be scrutinized. A burning question now is the quantum oscillation experiments indicating the existence of Fermi pockets in the putative normal state as opposed to the angle resolved photoemission spectroscopy (ARPES)~\cite{Damascelli:2003} exhibiting so-called Fermi arcs~\cite{Norman:1998}.
\item Complexity of the phase diagram. The so-called unconventional superconductors, of which high-$T_{c}$ cuprates are examples, are distinguished from their conventional cousins by their complex phase diagrams with many competing ground states with
quantum phase transitions between them. A complex phase diagram implies competing interactions in the reduced Hamiltonian, which in turn imply broken symmetries. It is remarkable that even Buckyball superconductivity that was once declared to be a conventional electron-phonon superconductor~\cite{Gunnarsson:1997}, with one serious exception~\cite{Chakravarty:1991}, is now found to be unconventional in the most recent experiments~\cite{Takabayashi:2009}---compare the phase diagram  of $\mathrm{Cs_{3}C_{60}}$ with A15 structure and a phase diagram of cuprates in Fig.~\ref{fig:DDW}. The complexity of the diagrams and the multiplicity of phases is evident. 
\begin{figure}[htb]
\begin{center}
\includegraphics[width=\linewidth]{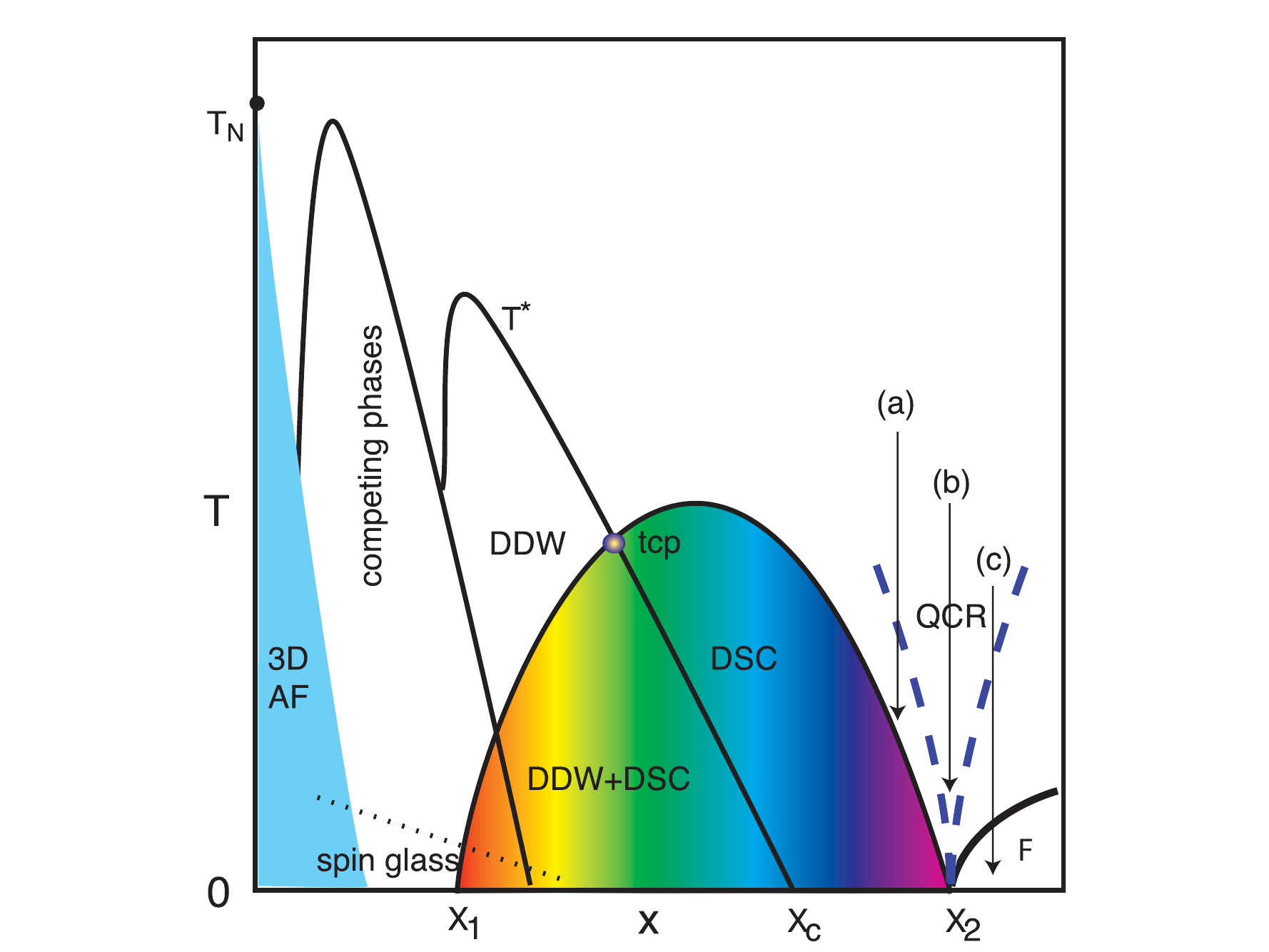}
\end{center}
\label{fig:DDW}
\caption[DDW]{\label{fig:DDW} There are many proposed phase diagrams for high-$T_{c}$ cuprates with a multiplicity of phases. Here is my favorite  one,  as a function of doping $x$, with emphasis on broken symmetries. The symbols are: 3D AF---three dimensional antiferromagnet; DDW is $d$-density wave; DSC is $d$-wave superconductor; $tcp$ stands for a tetracritical point; $x_{2}$ stands for a suggested quantum critical point (QCP) in the overdoped regime separating superconductivity from ferromagnetism~\cite{Kopp:2007}, F,  and $x_{1}$ and $x_{c}$ are two other QCPs. QCR is a depiction of the quantum critical region, and (a), (b), and (c) are possible experimental trajectories that should exhibit different behaviors as a function of temperature. The region termed competing phases may be composed of a cascade of various charge ordered states.}
\end{figure}
\end{itemize}
\section{Quantum oscillation experiments: a brief overview}
The purpose of these experiments is to crush the superconducting dome by applying a high magnetic field and to study the resulting normal state at low temperatures. After all, it is essential to understand the normal ground state if we are to get to the secret of high transition temperatures. The field range has been generally very high, $35-65$ T,  and in some cases even higher, up to 85 T. Nonetheless, as far as the energetics are concerned this is still very small compared to the electronic energy scale~\cite{Nguyen:2002}.
There have been  de Haas-van Alphen, Shubnikov-de Haas, and  Hall measurements. The materials are hole doped $\mathrm{YBa_{2}Cu_{3}O_{6+\delta}}$ (YBCO)  close to 10\% doping, $\mathrm{YBa_{2}Cu_{4}O_{8}}$ (Y124) corresponding to 14\% doping and overdoped single layer $\mathrm{Tl_{2}Ba_{2}CuO_{6+\delta}}$ (Tl2201). Further doping levels of YBCO have also been explored~\cite{Singleton:2009}. More recently Shubnikov-de Haas oscillations in the $c$-axis conductivity  have also been observed in optimal to overdoped  electron-doped $\mathrm{Nd_{2-x}Ce_{x}CuO_{4}}$ (NCCO) at 15\%, 16\%, and 17\% doping~\cite{Helm:2009}. This is significant for at least two reasons: (1) the material involves substantial cation disorder and (2) the measurements are carried out in the field range $35 - 65$ T,  far above $H_{c2}$ that is less than 10 T. The field range implies that in NCCO there is hardly any fluctuating order to speak off. The similarity of the observations with the hole doped cuprates is so striking that it is unlikely that mechanisms in electron and hole doped materials are different, {\em ipso facto} pairing fluctuations could not possibly play a major role in these experiments~\cite{Eun:2009}. There has been some controversy as to whether or not the quantum oscillation measurements in YBCO are carried out in the ``normal state''~\cite{Li:2007}, although it is certain that they are carried out in the resistive state.

In this brief space I can only highlight some striking features of the relevant experiments. I shall return at the very end to some of the unresolved puzzles.
\begin{itemize}
\item The first experiment detected oscillations of the Hall coefficient~\cite{Doiron-Leyraud:2007}. This is dramatic because within Fermi liquid and Boltzmann transport theories the scattering rate cancels out in the Hall coefficient as long as there is only one type of pocket, electron or hole. At least two types of pockets with differing mobilities are necessary to observe oscillations of the Hall coefficient~\cite{Chakravarty:2008b}, let alone its negative sign in a hole doped material. The existence of oscillation in the Hall coefficient is therefore indicative of two types of charge carriers.
\item The most plausible explanations of these oscillations in my opinion have involved reconstruction of the Fermi surface involving a density wave order~\cite{Chakravarty:2008b,Millis:2007}. Whether or not the order is commensurate or incommensurate is important but not the primary issue. 
\item Measurements in a tilted field have yielded conflicting results. In one case no interference effects leading to spin zeros in the oscillation amplitudes were reported~\cite{Sebastian:2009} and in the other spin zeros were found~\cite{Ramshaw:2010}. The authors of the latter experiment ascribe the difference to greater resolution in separating two close frequencies that have spin zeros at two different angles. One of them may reach the zero at an angle where the other is finite. Thus if these two frequencies are not resolved, it would appear that there are no spin zeros. Excessive tilting of the field may drive the the system deep into the superconducting state that possibly cannot exhibit quantum oscillations. While the normal component couples to the supercurrent, the Zeeman coupling is to the full field. These experiments are terribly significant: the existence of spin zeros implies that  the quasiparticles carry charge $e$, spin-$1/2$, a $g$-factor close to 2, and an effective mass $m^{*}$  within a factor of 2 of the free electron mass in underdoped cuprates---shocking after some three decades of complex Mott physics. One can go further and provide strong argument~\cite{Garcia:2010} that whatever density wave order is causing the Fermi surface reconstruction has to be of the singlet variety. Conversely, non-existence of spin zeros implies that the order parameter is likely to be a spin triplet.
\end{itemize}
\section{Broken symmetries}
A fundamental organizing principle of matter is the notion of broken symmetries. When a symmetry is present in the Hamiltonian but not in the ground state, we say that the symmetry is broken. For example, a liquid has all the symmetries of the Hamiltonian, rotation, translation, etc. In contrast a crystal has only a set of discrete translational and rotational symmetries. This definition is found to be counterintuitive by an artist who typically finds no symmetries of a featureless liquid but beautiful symmetries in a crystalline state or in an isicle. Broken symmetries often have their key signatures, if those are not hidden from our common sense experience. For example, a magnet breaks time reversal and spin rotational symmetries, a crystal symmetries of the space group, ferroelectrics the inversion, and superconductors the global gauge symmetry. Nonetheless, the roster of observed broken symmetries are far too small compared to the multitude of possibilities. In fact, it is a truism that any symmetry that can be broken, must be broken. So why are so  few broken symmetries observed in Nature? I would like to argue that in high temperature superconductors certain broken symmetries and the consequent order parameters are hidden from the common observational techniques. This does not imply that these hidden orders  are unimportant, but only that we do not have the tool to observe them. They work behind the scene.

Broken symmetries offer protection from irrelevancies. Deep inside a broken symmetry state elementary excitations are uniquely known.  This, in turn, allows us to understand and predict the  properties matter. Often mean field theory and small fluctuations about it are sufficient theoretical tools. Deep inside a broken symmetry phase critical fluctuations are absent and even  the collective modes can be physically identified from the symmetries of the order parameters. Moreover, those properties that are determined by symmetries can be determined in the weak interaction limit, where the calculations are better controlled, from a suitable effective Hamiltonian. The results thus obtained  should be at least qualitatively valid even in the strong interaction limit.~\cite{Kopp:2007} However, the thermal properties in low dimensional systems are badly predicted by the Hartree-Fock theories of broken symmetries, because fluctuations are often very important. Nonetheless, it is expected that the results at zero temperature retain a considerable degree of validity, except close to quantum critical points where quantum fluctuations become important. This leads us to the discussion of what  dominant symmetries are broken  in unconventional superconductors. 

\subsection{The density wave state of higher angular momentum: an example with application}
As there are excellent review articles of certain classes of broken symmetry phases~\cite{Kivelson:2003,Sachdev:2003}, I will concentrate here on a class that  I favor, about which there are no review articles. Additionally, understanding of these states may be important in understanding the quantum oscillation experiments, the major focus of the present article.
A whole class of density wave states relevant to cuprates can be defined by the angular momentum quantum number and the fundamental nature of the condensates~\cite{Nayak:2000}. A superconductor is a condensate of Cooper pairs, that is, the condensation is in the particle-particle channel. Thus, the overall antisymmetry of the wave function results in strict restrictions.  If the orbital function is symmetric, the spin function must be antisymmetric and vice versa. In contrast, the density wave states are condensates of bound pairs of electrons and holes. Since there is no requirement of exchange between the two distinct particles, the orbital function cannot constrain the spin function. Although angular momentum is not a strict quantum number in a crystal, we will continue to use it as a metaphor---the proper classification is in terms of the symmetry of the point group.

\begin{figure}[htbp]
\begin{center}
\includegraphics[scale=0.5]{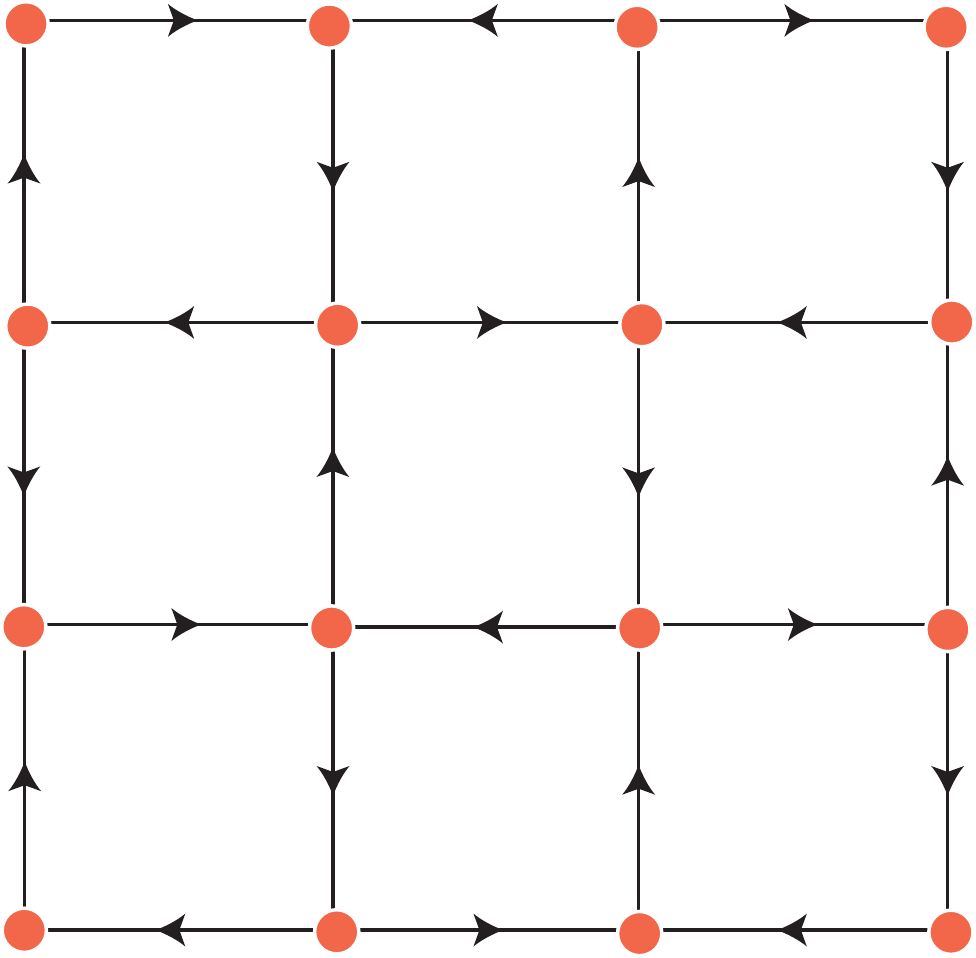}
\caption{Staggered pattern of charge currents reflecting DDW order in the square planar CuO-lattice with O-sites ommitted.}
\label{fig:ddw-current}
\end{center}
\end{figure}

For a superconductor $\ell = 0, 1, 2, \ldots$ define $s$-wave (spin singlet), $p$-wave (spin triplet), and $d$-wave (spin singlet) condensates, etc. For a particle-hole condensate, that is, a density wave, $\ell =0$ comes in two varieties, a spin singlet version and a spin triplet version. The spin singlet version is the familiar charge density wave (CDW) and the triplet version the spin density wave (SDW). The $\ell =1$ comes also in two versions and involves bond order. The focus of this article is the case $\ell=2$.  Here the spin singlet version is not a wave of density at all but corresponds to a staggered pattern of circulating charge currents, dubbed the $d$-density wave (DDW). The $\ell=2$ spin triplet version corresponds to a staggered pattern of circulating spin currents. The two-fold commensurate DDW breaks translation,  time reversal, parity, and a rotation by $\pi/2$, while the product any two symmetries is preserved. More specifically, the DDW order parameter is defined by ($a$ being the lattice constant)
\begin{equation}
\langle {c^{\alpha\dagger}}({\bf k}+{\bf Q},t)
{c_\beta}({\bf k},t)\rangle
= i\frac{\Phi_{\bf Q}}{2}\,(\cos{k_x}a-\cos{k_y}a)\, {\delta^\alpha_\beta},
\end{equation}
where ${\bf Q} = (\pi/a, \pi/a)$. Note the similarity of the form factor with the $d$-wave superconductor (DSC) and the crucial factor of $i$ signifying the breaking of time reversal symmetry. The Kronecker $\delta_{\beta}^{\alpha}$ reflects the fact that the order parameter transforms as identity in the spin space, hence a singlet. As mentioned above  the order parameter in the real space corresponds to a staggered pattern of circulating charge currents shown in Fig.~\ref{fig:ddw-current}.
It has been proposed~\cite{Chakravarty:2001} that the DDW gap is proportional to pseudogap $T^{*}$; see Fig.~\ref{fig:DDW}. From this proposed identification one can estimate that the largest magnetic field corresponding to current circulating around a square plaquette is about $10-30$ Gauss. A triplet DDW order on the other hand is given by 
\begin{equation}
\langle {c^{\alpha\dagger}}({\bf k}+{\bf Q},t)
{c_\beta}({\bf k},t)\rangle
= i\frac{\Phi_{\bf Q}}{2}\,(\cos{k_x}a-\cos{k_y}a)\, \hat{n}\cdot \boldsymbol{\sigma}^{\alpha}_{\beta},
\end{equation}
which clearly transforms as a triplet under rotation in spin space. In the real space this order parameter corresponds to circulating staggered spin currents. The unit vector $\hat{n}$ defines the direction of the spin.
\subsection{Hidden order}
The singlet or the triplet DDW order are  hidden from most common probes because they do not result in a net charge modulation nor a spin modulation. The reason is that
\begin{equation}
\sum_{{\bf k}\in BZ}\langle {c^{\alpha\dagger}}({\bf k}+{\bf Q},t)
{c_\beta}({\bf k},t)\rangle = 0.
\end{equation}
Contrast it with a d-wave superconductor, where the Meissner effect follows directly from the broken $U(1)$ symmetry, irrespective of the pairing channel. Thus, even though the integral over the Brillouin zone is zero, the superconducting order can be detected by an applied uniform magnetic field $B$ through the Meissner effect. We have no such luxury in the case of DDW. Even specific heat is smooth as a function of temperature, unlike a superconductor~\cite{Chakravarty:2002}. Experiments seeking to uncover such order must (a) be sensitive to spatial variation of the kinetic energy or currents, (b) measure higher order correlations of charge or spin densities (2-magnon Raman, NQR, etc.). For triplet DDW, we have broken  spin rotational symmetry and the corresponding Goldstone mode, which can be picked up in neutron scattering measurements, as can the staggered magnetic field for the singlet DDW.  A triplet DDW does not even break time reversal and so it is hidden further. The effect of potential disorder is rather subtle in the DDW state; it is rather robust in comparison to a $d$-wave superconductor (DSC)~\cite{Ghosal:2004}. This legitimizes the use of a mean field Hamiltonian with a given order parameter along with disorder without performing a self consistent determination of the order parameter. The reason, in essence, is that DDW involves current modulation as opposed to density modulation. 

\section{What is  and what is not a Fermi liquid?}

There is a great schism: a class of experiments undoubtedly point to non-Fermi liquid behavior and a class of experiments that tantalizingly resemble, even quantitatively to some degree, a Fermi liquid normal state at low temperatures when the superconducting dome is crushed by a high magnetic field, namely the quantum oscillation measurements. The distinction between a Fermi and a non-Fermi liquid is sharpest in the ground state, not at finite temperatures. From this perspective it does not matter whether or not there are sizable superconducting fluctuations above $T_{c}$, which I believe is perfectly reasonable, especially for superconductors with high pairing scale.
The most convincing experiment that points to a Fermi liquid is a recent Shubnikov-de Haas experiment involving multiple oscillation frequencies~\cite{Ramshaw:2010} that are interpreted as conventional quantum oscillations as though the quasiparticles have charge $e$, spin $1/2$, and a $g$-factor close to $2$, more precisely $2.2$.  In order to appreciate this issue, we need to take a step back and discuss what is and what is not a Fermi liquid. Since much is known about Fermi liquids, I will concentrate on what is not a Fermi liquid.

Only non-analyticity of the 
time-ordered one-particle Green's function on the principal
sheet, in the infinite volume limit,   is a cut along the real axis. To describe a Fermi liquid, Galitskii and
Migdal\cite{Galitskii:1958} proposed a model of the one-particle Green's function, which
has  simple poles, sufficiently close to the real axis in the second and the fourth quadrants on the second sheet.
The pole in the  fourth quadrant
corresponds to a long-lived quasiparticle and that in the second quadrant to a long-lived quasihole.
According to this model 
the matrix elements of a fermion operator between the ground state and the exact
eigenstates is sharply peaked at an energy depending  on the
wavevector {\bf k}. The quasiparticle pole in the complex plane is the representation
of the physical phenomena on the real frequency axis, as the theory of analytic
continuation implies. A simple pole  implies a unique energy
 $\varepsilon_{\bf k}$ for a given ${\bf k}$. With proper normalization of the wave function, a
quasiparticle can be made to resemble a bare particle for energies asymptotically
close to the Fermi energy. Other properties of quasiparticles also follow from the
assumptions of the model.  The charge of the quasiparticle is  $e$  and its spin  $1/2$.

The Galitskii-Migdal model can be summarized by the spectral function
\begin{equation}
A({\bf k},\omega)= -{1\over \pi}{\rm Im}\; G_{\rm R}({\bf k},\omega) \nonumber =z_k\delta(\omega-(\varepsilon({\bf k})-\mu)),
\end{equation}
where $G_{\rm R}$ is the retarded Green's function, $z_k$ is the
quasiparticle residue and $\mu$ is the interacting chemical potential. As this model is  valid only for quasiparticles
close to the Fermi surface, we can linearize:  $\varepsilon
({\bf k})-\mu\approx v_{{\bf k}^{*}}|{\bf k}-{\bf k}_{*}|$, where
${\bf k}_{*}$ is the point on the  Fermi surface nearest to ${\bf k}$; for a Fermi surface with spherical symmetry $k_{*}=k_{F} $, the Fermi wave vector, and $v_{{\bf k}^{*}}=v_F$, the Fermi velocity. Then, the asymptotic spectral
function has the scaling  property
\begin{equation}
A(\Lambda |{\bf k}-{\bf k}_{*}|,\Lambda\omega)=\Lambda^{-1}A(|{\bf k}-{\bf k}_{*}|,\omega),
\end{equation}
reflecting the gaplessness of  the elementary
excitations. The system is poised at criticality. That a
normal Fermi system represents a critical system is, of course, well known, but
usually this  criticality  does not lead to
serious dynamical consequences. This is  because the modes (the fermionic excitations)
decouple, as in the Gaussian model of  classical critical phenomena.  An exception is the Kondo problem where the criticality leads  to non-trivial coupling between the modes, resulting in a breakdown of the single particle picture~\cite{Wilson:1983}.

What would be the corresponding model of a non-Fermi liquid?  A well defined model  could be one where  the analytic
continuation of the Green's function to the second sheet have branch points instead
of simple poles~\cite{Yin:1996}. Thus, the corresponding scaling relation for the asymptotic spectral function is
\begin{equation}
A(\Lambda^{y_1} |{\bf k}-{\bf k}_{*}|,\Lambda^{y_2}\omega)= \Lambda^{y_A}A( |{\bf k}-{\bf k}_{*}|,\omega),
\end{equation}
where $y_1$, $y_2$, and $y_A$ are the  exponents defining the universality class of
the critical Fermi system. The exponents other than 
$y_1=1$, $y_2=1$, and $y_A=-1$ 
have been termed anomalous.  The notion of a spectral
function with anomalous exponents was termed {\em spectral anomaly}. 
Spectral anomaly  implies that a given $\bf k$ does not correspond to
a single frequency but a continuum of frequencies. A creation operator of wave vector
{\bf k}, applied to the ground state, creates a state that couples  to many
eigenstates of the unperturbed system that can no longer be considered a damped quasiparticle.
The  analytic property of the Green's function is completely changed.

Continuity with the Fermi liquid model requires that the location of the branch
point  be in one-to-one correspondence with the poles. This
is by no means an obvious assumption, but holds for one-dimensional electron gas
~\cite{Mattis:1965}, and  also  for  gauge models\cite{Nayak:1994a,Nayak:1994b} in higher dimensions.
We have not incorporated spin-charge separation in our
definition. At the level of generality stated above the inclusion of fractionalization  is straightforward, but simply more 
elaborate.

From the dispersion relation
we can now determine the real part of the retarded Green's function, i.e,
\begin{equation}
{\rm Re}G_{\rm R}(|{\bf k}-{\bf k}_{*}|,\omega)=-{{\rm P}\over \pi}\int_{-\infty}^{+\infty}
d\omega'{{\rm Im}G_{\rm R}(|{\bf k}-{\bf k}_{*}|,\omega)\over \omega - \omega'}.
\end{equation}
Thus, the real part also satisfies the same
scaling relation given by
\begin{equation}
{\rm Re}G_{\rm R}(\Lambda^{y_1||\bf k}-{\bf k}_{*}|,\Lambda^{y_2}\omega)=
\Lambda^{y_A}
{\rm Re}G_{\rm R}(|{\bf k}-{\bf k}_{*}|,\omega).
\end{equation}
From the generalized homogenity assumption, we also get
\begin{equation}
A(|{\bf k}-{\bf k}_{*}|,0)={1\over |{\bf k}-{\bf k}_{*}|^{y_A/y_1}}A(1,0)
\end{equation}
and 
\begin{equation}
A(0,\omega)={1\over |\omega|^{y_A/y_2}}A(0,1).
\end{equation}
The momentum distribution function $n(k)$ is given by
\begin{equation}
n(|{\bf k}-{\bf k}_{*}|)=\int_{-\infty}^0 d\omega A(|{\bf k}-{\bf k}_{*}|,\omega).
\end{equation}
Using the scaling relation, we can write
\begin{equation}
n(|{\bf k}-{\bf k}_{*}|)=\int_{-\infty}^0 d\omega
\Lambda^{-y_A}A(\Lambda^{y_1}[{\bf k}-{\bf k}_{*}],\Lambda^{y_2}\omega).
\end{equation}
Because $\Lambda$ is an arbitrary scale factor, we can chose
$\Lambda^{y_1}|{\bf k}-{\bf k}_*|=1$. Then, 
\begin{equation}
n(|{\bf k}-{\bf k}_{*}|)=|{\bf k}-{\bf k}_{*}|^{(y_2+y_A)/y_1}\int_{-\infty}^0 dx A(\pm 1,x).
\end{equation}
The  discontinuity present in a 
Fermi lquid  is destroyed due to spectral anomaly. In a Fermi system, the occupation
of a state cannot  diverge, and the
inequality 
$
(y_2+y_A)/ y_1 > 0 
$
must be satisfied. There is a superficial difficulty with the scaling argument above:
the integral does not converge. This is a minor point and is true for any leading asymptotic 
scaling function; we need to restore the cutoff to correct it.
However, the critical exponent is still given correctly by the above
scaling argument. 

Fermi liquid theory is an
effective low energy theory of
metals\cite{Polchinski:1992,Shankar:1994}. We shall briefly 
recapitulate the arguments due to Polchinski\cite{Polchinski:1992}. Imagine that we
are considering an effective field theory below an energy scale
$E_c$ at which there are strong Coulomb interactions. 
Consider first the free action (the repeated spin index $\sigma$ is summed over):
\begin{equation}
\int dt d^d{\bf k}\left\{ i\psi^{\dagger}_{{\bf
k}\sigma}(t)\partial_t \psi_{{\bf
k}\sigma}(t)-\left(\varepsilon({\bf k})-\mu\right)\psi^{\dagger}_{{\bf
k}\sigma}(t)\psi_{{\bf
k}\sigma}(t)\right\}
\end{equation}
As we scale all energies by a factor $s < 1$, the momenta must scale to
the Fermi surface. To accomplish this scaling, we write 
\begin{equation}
{\bf k}={\bf k}_{*}+{\bf l}.
\end{equation}
 As we scale the energies and the wave
vectors,
$E\to sE$,
${\bf k}_{*} \to {\bf k}_{*}$, and
${\bf l}\to s{\bf l}$, and the action remains fixed provided  the
dimension of the fermion operators is
$s^{-1/2}$.  If there are no relevant operators, the
effective low energy theory is self-consistent. The generated mass term 
can be absorbed by redefining the fermi surface to be  the true
interacting Fermi surface.

The most important interaction is the four fermion operator:
\begin{eqnarray}
\int dt&& d^{d-1}{{\bf k}_{*1}}\; d{{\bf l}_1}\; d^{d-1}{{\bf k}_{*2}}\; d{{\bf l}_2}\; 
d^{d-1}{{\bf k}_{*3}}\;  d{\bf l_3} \; d^{d-1}{{\bf k}_{*4}}\; d{\bf l_4} V({\bf k}_{*1},
{\bf k}_{*2},{\bf k}_{*3}, {\bf k}_{*4})\nonumber \\
&&\psi^{\dagger}_{{\bf k}_1\sigma}(t)\psi_{{\bf k}_3\sigma}(t)
\psi^{\dagger}_{{\bf k}_2\sigma'}(t)\psi_{{\bf k}_4\sigma'}(t)
\delta^d({\bf k}_1+{\bf k}_2-{\bf k}_3-{\bf k}_4),
\end{eqnarray}
where we have assumed that the interaction is short-ranged. It can be seen that
the interaction scales as
$s$ times the  dimension of the delta function.  Let us assume, for the moment, that
\begin{eqnarray}
\delta^d({\bf k}_1+{\bf k}_2-{\bf k}_3-{\bf k}_4)&=&
\delta^d({\bf k}_{*1}+{\bf k}_{*2}-{\bf k}_{*3}-{\bf k}_{*4}+{\bf l}_1+{\bf l}_2-{\bf
l}_3-{\bf l}_4) \nonumber \\
&\sim&\delta^d({\bf k}_{*1}+{\bf k}_{*2}-{\bf k}_{*3}-{\bf k}_{*4}),
\end{eqnarray}
where we have ignored $\bf l$ in comparison to ${\bf k}_{*}$,  because 
$l$ all scale to zero. Now the argument of the $\delta$-function does not depend on
$s$, and the four fermion interaction scales as $s$, vanishing in the limit $s\to
0$. The interaction is irrelevant! However, the argument fails for
the special kinematics implied by the forward, the exchange, and the Cooper channel
scattering processes. For these processes, the delta function can be seen to be of
order
$s^{-1}$, transforming the dimension of the interaction  to $s^0$. The net result
is that the interaction is marginal\cite{Polchinski:1992}. 
The argument also  fails in $d=1$ because the $\delta$-function is always of order $s^{-1}$ and the
interaction is always marginal. 

We now turn to the consistency of our non-Fermi liquid model~\cite{Yin:1996}. Let the action be 
\begin{equation}
\int d\omega\;  d^d{\bf k}\;  G^{-1}({\bf k},\omega)\psi^{\dagger}_{{\bf k}\sigma}(\omega)
\psi_{{\bf k}\sigma}(\omega),
\end{equation}
where the Green's function $G$ corresponds to that of a non-Fermi liquid defined above. The dimension
of  $\psi_{\sigma}({\bf k},t)$ is $s^{y_{A}/2}$. If we follow 
Polchinski, we see that the  the four fermion
interaction is irrelevant; even for those exceptional kinematics for which the
interaction was marginal in the Fermi liquid case, it now scales as $s^{2(1+y_{A})}$, hence irrelevant as long as there is spectral anomaly, that is, $(1+y_{A})>0$. A non-Fermi liquid 
 is more stable than a Fermi liquid! In fact, in the weak coupling
regime, it does not even allow a superconducting instability. The coupling has to reach a
threshold  before superconducting instability occurs. This is a remarkable result. {\em In a Fermi liquid-BCS theory, arbitrarily weak attractive interaction leads to a superconducting instability, hence there is no chance of a quantum critical point, as long as the interaction is attractive. In contrast, in a non-Fermi liquid there can be a quantum critical point because the coupling has to reach a finite threshold before superconductivity sets in.}

\section{Quantum oscillations: Kohn's theorem}

The periodicity observed in all quantum oscillations has its root in the Landau levels. There is an instructive theorem due to Kohn~\cite{Kohn:1961} that exemplifies the robustness of the effect. Consider a continuum two-dimensional electronic system with arbitrary short ranged  velocity independent electron-electron interaction, but without disorder. The Hamiltonian is
\begin{equation}
{\cal H} = \int_{\mathbf x} \psi_{\sigma}^{\dagger}({\mathbf x})\varepsilon({\mathbf p}, {\mathbf x}) \psi_{\sigma}({\mathbf x}) + \int_{{\mathbf x},{\mathbf x'}}\psi_{\sigma}^{\dagger}({\mathbf x})\psi_{\sigma'}^{\dagger}({\mathbf x'}) v({\mathbf x}- {\mathbf x'})\psi_{\sigma'}({\mathbf x'})\psi_{\sigma}({\mathbf x})
\end{equation}
Here $\varepsilon({\mathbf x},{\mathbf p})$ is the Hamiltonian for individual particles of mass $m$, charge $e$, magnetic moment $\mu_{B}$, spin $\frac{1}{2}$, moving in an external magnetic field $\mathbf B$. Here $v({\mathbf x}- {\mathbf x'})$ is assumed to be a spin independent static potential, so any explicit spin dependence can be dropped. Moreover, for simplicity we shall also drop the Zeeman term $-g\mu_{B}{\mathbf B}\cdot \boldsymbol{\sigma}$ in the Hamiltonian. The single particle energy is 
\begin{equation}
\varepsilon({\mathbf p},{\mathbf x}) = \frac{1}{2m}\left[ {\mathbf p} - (e/c) {\mathbf A}\right]^{2}.
\end{equation}
 We choose the Landau gauge ${\mathbf A} = (0, -Hx, 0)$. Note that since magnetic field is {\em never} a small perturbation, as it increases indefinitely  with the size of system, it is  sensible  to begin with the noninteracting system but in the presence of the magnetic field and then consider the effect of the interaction term. 
 
The solution to the single particle problem  is the well-known Landau levels with  energy eigenvalues and eigenfunctions  given by 
\begin{equation}
\epsilon_{n,{\mathbf k}}=\hbar\omega_{c}\left(n+\frac{1}{2}\right), \; \psi_{n, k}= e^{iky}u_{n}\left(x+\frac{\hbar c k}{eB}\right),
\end{equation}
where the frequency $\omega_{c}=\frac{eB}{mc}$ and the degeneracy of each energy level is
\begin{equation}
d_{\Phi}=2\frac{\Phi}{\Phi_{0}} ,
\end{equation}
where  $\Phi= B L_{x}L_{y}$ and  $\Phi_{0}=hc/e$; the factor of 2 is for spin. The functions $u_{n}$ are the one-dimensional harmonic oscillator wave functions.

In a magnetic field momenta are not good quantum numbers, but it is still useful to depict the energy levels on  the two-dimensional $k_{x}-k_{y}$-plane. The spectra, of degeneracy $d_{\Phi}$, lie on concentric circles in this plane separated by $\hbar\omega_{c}$.  The total number of available states  in an  area $\Delta A=\frac{2\pi m}{\hbar^{2}}\delta\varepsilon$ between the two energy surfaces  is unchanged despite the Landau quantization:
\begin{equation}
\frac{L_{x}L_{y}}{(2\pi)^{2}}\Delta A=\frac{d_{\Phi}}{2}.
\label{eq:degeneracy}
\end{equation}
Let the Fermi level, $\epsilon_{F}$, at $T=0$ be situated on one such concentric level. Then  all states with energy $E\le  \epsilon_{F}$ are completely  filled and all states for $E> \epsilon_{F}$ are completely empty.  The area enclosed by the Fermi level, $A(\epsilon_{F})$, follows trivially from Eq.~(\ref{eq:degeneracy}) and is
\begin{equation}
2\frac{A(\epsilon_{F})}{(2\pi)^{2}} = \frac{N}{L_{x}L_{y}},
\end{equation}
where $N$ is the {\em fixed}  total number of particles including both spin directions. That is none other than the Luttinger sum rule~\cite{Luttinger:1960}, relating the volume of the Fermi surface and the density of particles despite the  magnetic field. 

The magnetic field  with  an integer number, $j$, of the lowest Landau levels completely filled and the rest empty will satisfy
\begin{equation}
2\frac{A(\epsilon_{F})}{(2\pi)^{2}} = \frac{1}{L_{x}L_{y}} \frac{2\Phi}{\Phi_{0}} j
\end{equation}
or
\begin{equation}
\frac{1}{B_{j}}= j\frac{2\pi e}{\hbar c}\frac{1}{A(\epsilon_{F})}.
\label{eq:Bn}
\end{equation}
This isolated nondegenerate ground state is  separated by a gap $\hbar \omega_{c}$ from the excited state. As we increase $B$, the quantized orbits are drawn out of the Fermi level, and sequentially pass through essentially~\footnote{This is certainly true as long as we are not at a very low Landau level. Otherwise, the difference in the free energy between the $n$-th filled Landau level and $(n-1)$-th filled Landau level is substantial and cannot be dropped as a small correction.} identical set of nondgenerate isolated ground states, resulting  in the periodicity of the properties of the electron gas. 
The frequency, $F$, of the oscillations when analyzed in terms of $1/B$ is then the universal relationship
\begin{equation}
F= \frac{\hbar c}{2\pi e} A(\epsilon_{F}).
\end{equation}
Periodicity of course does not imply a sinusoidal wave form.

The interesting question now is what we can say about the interacting systems.
Now, fix $B_{j}$ to a completely filled Landau level and turn on the electron-electron interaction adiabatically. This is clearly possible for this incompressible state because there is a gap in the spectrum: $\hbar \omega_{c}$. Then to all orders in perturbation theory an isolated nondegenerate ground state will remain so. Therefore the sequence of states corresponding to fully filled Landau levels as a function of the magnetic field will be the same, as in the noninteracting case. The periodicity is therefore unchanged and is determined by the enclosed area $A(\epsilon_{F})$, which in turn is fixed by the Luttinger sum rule.~\footnote{Although we merely made use of continuity, the true interacting state may not be unitarily equivalent to the noninteracting state --- typically when a symmetry is broken.  For a partially filled  Landau level, which is degenerate, there is a zoo of density wave states for higher Landau levels~\cite{Koulakov:1996}.} Note that the theorem makes no statement about the amplitude of the oscillations, nor about the waveform of the periodicity.

It appears that we have only made use of the adiabatic theorem and as such Kohn's theorem may be true  non-perturbatively and perhaps can be generalized to a non-fermi liquid state. Despite motivated effort the present author was unable to generalize this theorem.~\footnote{This is not simply a consequence of Gell-Mann and Low's theorem~\cite{Gell-Mann:1951} applied to a non-degenerate ground state with a gap such that the adiabatic theorem can be applied. Although for Fermi liquids adiabatic switching evolves free particles to quasiparticles, one must guard against symmetry change and unitary inequivalence, or even the difference in the analytic structure of the Green's function of a non-Fermi liquid versus a Fermi liquid.} The real question is do non-Fermi liquids have Landau levels? Perhaps a phenomenological illustration is useful.
For a Fermi liquid, the quasiparticle spectral function is a series of $\delta$-functions, which is trivially (considering for illustration for the $2D$ case)
\begin{equation}
A(\epsilon) =2 \frac{eBL_{x}L_{y}}{2 \pi\hbar c}\sum_{n}\delta[\epsilon - (n+1/2)\hbar \omega_{c}]
\end{equation}
with appropriate renormalization of the effective mass. Considering that the ground state corresponds to all Landau levels filled up to an energy $\epsilon$ in the $k_{x}-k_{y}$-plane denoted by $\pi k_{F}^{2}$, as discussed above, we immediately obtain
\begin{equation}
\Delta\left(\frac{1}{B}\right) = \frac{2\pi e}{\hbar c}\frac{1}{A(k_{F})}.
\label{eq:Onsager}
\end{equation}
One might wonder if a generalization to a non-Fermi liquid might not exist. We could write
\begin{equation}
A_{NFL}(\epsilon)\sim \sum_{n}[\epsilon - (n+1/2)\hbar \omega_{c}]^{y_{A}/y_{2}}
\end{equation}
where the appropriate anomaly exponents were defined earlier; note that $y_{A}/y_{2}<0$. This will imply once again that the spectral functions have periodic branch points at the same values as in Eq.~\ref{eq:Onsager}. Despite its intutive appeal, it remains to be proven that a non-Fermi liquid has Landau levels.

\subsection{Luttinger formalism}
In an often quoted paper by Luttinger~\cite{Luttinger:1961} a derivation of the Lifshitz-Kosevich (LK)  formula~\cite{Abrikosov:1988} is given. But this formula is valid {\em if and only if} the Fermi liquid theory is valid. The modification of LK formula by Luttinger may include Fermi liquid corrections, but cannot be used for a non-Fermi liquid. It implies in turn that if the LK formula holds, we are very likely observing quasiparticles, perhaps with renormalized masses and with Fermi liquid corrections to the spin susceptibility if we are considering interference due to spins.

It is useful to recapitulate what Luttinger really proved. The range of interest is $k_{B}T, \; \hbar\omega_{c} \ll \mu$ and $k_{B}T$  not much larger than $\hbar\omega_{c}$. The self energy $\Sigma$ is separated into three parts:
\begin{equation}
\Sigma=\Sigma_{0}+\Sigma_{T}+\Sigma_{osc}.
\end{equation}
 Here $\Sigma_{0}$ is a {\em field independent} part taken at $T=0$ and $\Sigma_{T}$ is the first temperature correction, which by the Sommerfeld expansion is 
 \begin{equation}
 \Sigma_{T}\sim (k_{B}T/\mu)^{2}\Sigma_{0}.
 \end{equation}
Moreover  Luttinger estimates that
\begin{equation}
\Sigma_{osc}\sim (\hbar\omega_{c}/\mu)^{3/2},
\end{equation}
and therefore 
 \begin{equation}
 \Sigma_{osc}/\Sigma_{T}\sim (\hbar\omega_{c}/k_{B}T)^{2} (\mu/\hbar\omega_{c})^{1/2} \gg 1,
 \end{equation}
 allowing us to drop $\Sigma_{T}$.
 An assumption here is that the system is three dimensional and the Sommerfeld expansion is a meaningful asymptotic expansion. These estimates allow him to drop $\Sigma_{T}$. If we now assume that the electron-electron scattering rate  vanishes as $(\epsilon-\mu)^{2}$, which is a Fermi liquid assumption, the leading oscillatory part of the thermodynamic potential is
 \begin{equation}
 \Omega_{osc}= - \frac{1}{\beta} \sum_{r}\ln \left[ 1 +e^{\beta(\mu - E_{r})}\right],
 \end{equation}
 which is exactly the thermodynamic potential for independent fermions but with the renormalized quasiparticle energies $E_{r}$ determined from 
 \begin{equation}
E_{r} - Q_{r}(E_{r})=0,
\end{equation}
where $Q_{r}$ is the real part of the self energy.
 Except for that it is exactly the LK formula.
 
 The final formula of Luttinger can of course be cast in terms of fermionic  Matsubara frequencies, $\omega_{n}$,  and a self energy dependent on it. It then reads
 \begin{equation}
 \Omega_{osc}=-\frac{1}{\beta}\sum_{n} \mathrm{Tr}\{\ln\left[\epsilon+\Sigma_{0}(\omega_{n})-i\omega_{n}\right]\}.
 \end{equation}
 It appears that the non-oscillatory part of the self energy $\Sigma_{0}(\omega_{n})$  can be phenomenologically assigned  whatever we wish, in particular a non-Fermi liquid or a marginal Fermi liquid form. This procedure~\cite{Wasserman:1991,Pelzer:1991,McCollam:2008} seriously lacks consistency, as there is no proof that such systems exhibit Landau levels, which I believe is the {\em sine qua non} of quantum oscillations (cf. below).
 
\subsection{Disorder}

Any amount of impurities will break translational invariance and Kohn's theorem cannot, strictly speaking, hold. A moderate amount of disorder can be understood in terms of a self consistent Born approximation and has been discussed extensively~\cite{Ando:1974}. Landau levels will be broadened and will overlap, but as long as the states within a Landau band are not fully localized, magnetic oscillations persist. Can we understand, on dimensional grounds, how the oscillations are affected by disorder. Of course, impurity broadening must lead to the decay of the amplitudes characterized by the Dingle factors. How about the frequency? With disorder we have a new dimensionless parameter, $\hbar/(\epsilon_{F}\tau)$, that is expected to affect the frequency as well. The electron will take longer to complete a cyclotron orbit, so the frequency should be shifted downward, but by what amount? The downward shift in the energy $\epsilon$ of an extremal orbit can be estimated to be
\begin{equation}
\Delta(\epsilon) = \frac{P}{\pi}\int d\omega \frac{\Gamma(\epsilon,\omega)}{\epsilon-\omega},
\end{equation}
where $\Gamma(\epsilon_{k},\omega)=\pi \sum_{k\ne k'}|V_{k,k'}|^{2}\delta(\epsilon_{k'}-\omega)$ and $\epsilon_{k}=\epsilon=\epsilon_{F}$. When averaged over the distribution of disorder, $\Gamma$ is smooth and independent of energy, and therefore the principal value integral vanishes. A more refined self consistent argument~\cite{Goswami:2008} shows that the correction to the frequency is of order $(\hbar/\epsilon_{F}\tau)^{2}$, which is a small correction in most cases.

\section{The $\nu=1/2$ quantum Hall effect and transverse gauge field: an aside}
There are examples where quantum oscillations arise from a nearby non-Fermi liquid state. A remarkable example is the $\nu=1/2$ quantum Hall state. As the filling factor moves away from $\nu=1/2$, Shubnikov-de Haas oscillations are observed~\cite{Leadley:1994,Du:1994} and  the LK formula provides a good description, given  suitable redefinitions of the parameters of  a Landau Fermi liquid by the parameters corresponding to composite fermions~\cite{Jain:2007}. In particular, it is believed that the state at $\nu=1/2$ is a non-Fermi liquid with a logarithmically divergent effective mass due to an emergent transverse gauge field~\cite{Halperin:1993}. That transverse gauge fields are special in this respect and was first discovered by Holstein, Norton, and Pincus~\cite{Holstein:1973}. In fact, certain aspects of the de Haas-van Alphen oscillations were also noted by them, but the oscillations considered there were hardly affected by the non-Fermi liquid state because the gauge coupling constant was negligibly  small. In recent years emergent transverse gauge field in strongly correlated electron systems has been widely discussed~\cite{Wen:2004} where the effective coupling constant can be of order unity~\cite{Altshuler:1994,Chakravarty:1995}. It is an interesting question if  there may not be an underlying Fermi liquid (not necessarily a Landau Fermi liquid) responsible for the quantum oscillation experiments. That the normal state of high temperature superconductors are anomalous is well known, and it may be worthwhile to pursue this idea in some depth, if only to prove its invalidity.
 
\section{Fermi surface reconstruction: density waves}
\begin{figure}[bt]
\begin{center}
\includegraphics[width=\linewidth]{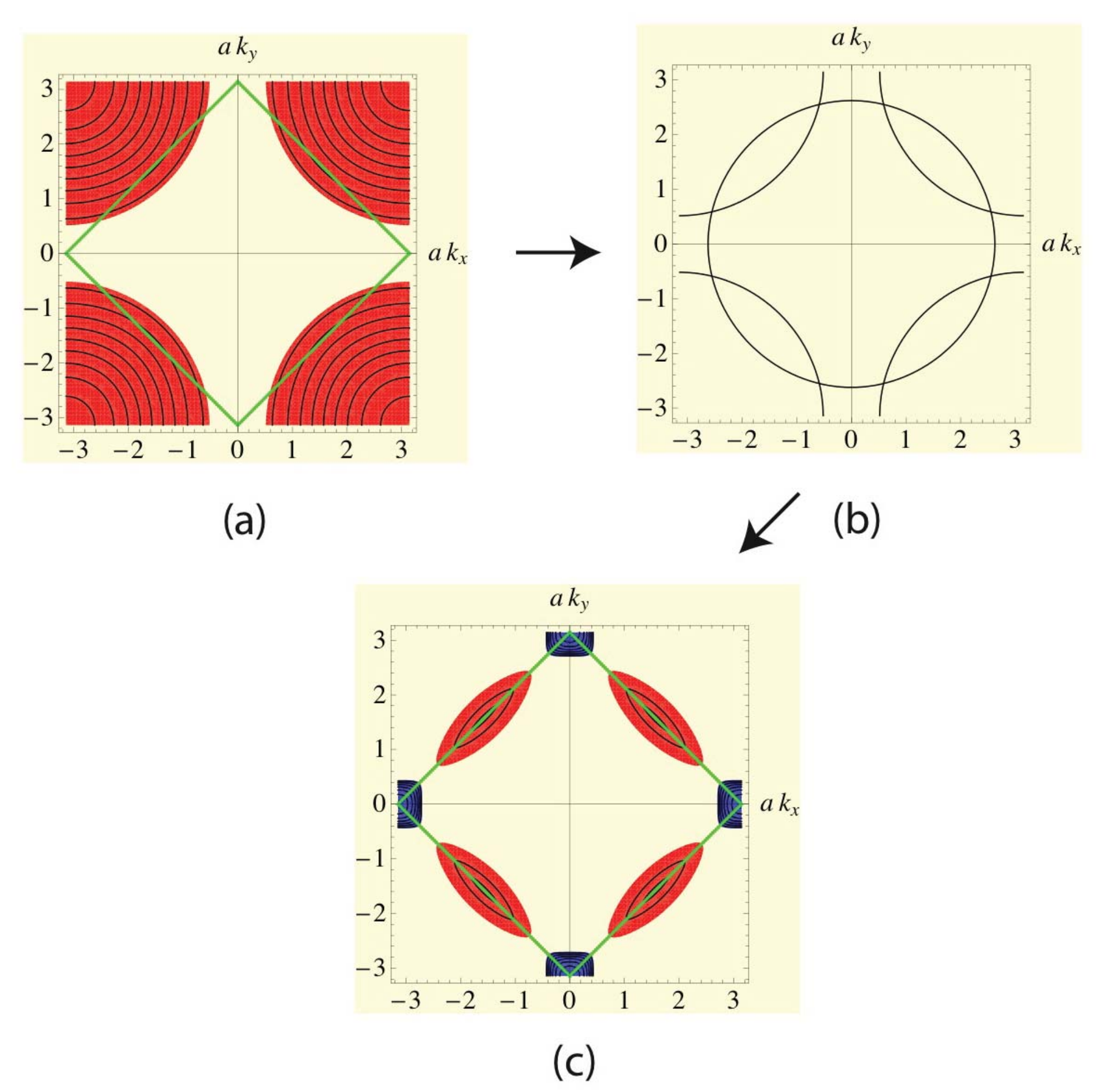}
\end{center}
\vspace*{8pt}
\caption{An illustration of a Fermi surface reconstruction, as described in the text. The red pockets represent holes and the blue pockets electrons. }
\label{fig:reconstruction}
\end{figure}

I will use the example of singlet  DDW to illustrate how Fermi surface reconstruction takes place, but any other two-fold commensurate order  at the mean field level will do. The Hamiltonian  in terms of the fermion creation and destruction  operators, $c^{\dagger}_{\bf k}$ and $c_{\bf k}$,  is (the spin index is ignored)
\begin{equation}
    H_1 = \sum_{{\bf k}\in RBZ}\left(\epsilon_{\bf k}c^\dagger_{\bf k}c_{\bf k}
        +\epsilon_{{\bf k}+{\bf Q}}c^\dagger_{{\bf k}+{\bf Q}}c_{{\bf k}+{\bf Q}}\right)
        +\sum_{{\bf k}\in RBZ}
        (iW_{\bf k}c^\dagger_{\bf k}c_{{\bf k}+{\bf Q}}+h.c.),
    \label{Eq:pureHamiltonian}
\end{equation}
where 
$\epsilon_{\bf k}$ is the single particle spectrun.  The reduced Brillouin
zone (RBZ) is bounded by $k_y \pm k_x = \pm\pi/a$. We define 
the DDW gap  by
\begin{equation}
W_{\bf k} = \frac{W_0}{2}(\cos k_x a - \cos k_y a),
\end{equation}
which is obviously proportional to the order parameter $\Phi_{\bf Q}$ defined earlier.
The resulting Fermi surface reconstruction is shown in Fig.~\ref{fig:reconstruction}. The diamond in ({\bf a}) is called the reduced Brillouin zone (RBZ) and contains exactly half the number of available states.  In ({\bf a}) the unoccupied states are colored red. The constant energy contours are shown as the set of black curves. The filled diamond corresponds to one electron per unit cell. In ({\bf a})  the red area corresponds to $(1+x)$ holes per unit cell of the crystal lattice. The excess, $x$, is called the doped holes.
 Consider shifting the Fermi surface in ({\bf a}) by vectors $(\pm \pi/a,\pm \pi/a)$, which will give rise to ({\bf b}), ignoring the shading for clarity. Matrix elements at the degeneracy points open up gaps, reconstructing the Fermi surface shown in ({\bf c}), as in a kaleidoscope. However, if we continue to consider the full BZ, we would double the number of states. All distinct states are contained  in the RBZ, but there are now two distinct sets of energy levels, the upper band and the lower band. However, we continue to use the full  BZ in ({\bf c}), as a better aid for visualization. Because the RBZ is the fundamental unit in the wave vector space, the new unit cell of the crystal lattice is doubled, given by a square $\sqrt{2} a\times\sqrt{2} a$, and the full translational symmetry of the original lattice is broken. The Fermi surface now consists of {\em disconnected sheets} of blue and red areas. The remarkable fact is that the charge carriers in the blue region behave like electrons of fraction $n_{e}$ and in the red region like holes  of fraction $n_{h}$. The doped holes are given by $x=2n_{h}-n_{e}$, as there are two hole pockets and one electron pocket in the RBZ from Luttinger sum rule (cf. below). The broken symmetry invoked here is called commensurate, strictly two-fold commensurate, as the translational invariance of the crystal of integer multiples of the next nearest neighbor  lattice vectors  of the original lattice is still preserved.

\subsection{Luttinger sum rule}

It would appear that broken symmetry  density wave states will lead to problems regarding the count of the charge carriers because the system is no longer in  one-to-one correspondence with free fermions. This is not the case~\cite{Altshuler:1998}. At the mean field level a whole class of density wave states with broken translational symmetry in the particle-hole channel, such as the spin density wave (SDW), the singlet and the triplet  $d$-density waves (DDW), are Fermi liquids in disguise when viewed in terms of quasiparticles of the valence and the conduction bands with their appropriate coherence factors. This mean field picture reconstructs the Fermi surface in terms of Fermi pockets, as illustrated for a commensurate DDW in Fig.~\ref{fig:reconstruction}. The full Hamiltonian is now the Hartree-Fock Hamiltonian plus the Hamiltonian of the residual interactions of the Hartree-Fock quasiparticles. Therefore, the proof of Luttinger's sum rule should be identical to that of a Fermi liquid~\cite{Luttinger:1960}. To understand when this will cease to hold, it might be useful recapitulate the conventional proof. The total number of particles $N$ can be written as
\begin{eqnarray}
N&=& - i \textrm{Tr}\sum_{\bf k}\lim_{t\to 0+}\int_{-\infty}^{+\infty}\frac{d\omega}{2\pi}G({\bf k},\omega)e^{i\omega t}\nonumber \\
   &=& i \textrm{Tr}\sum_{\bf k}\lim_{t\to 0+}\int_{-\infty}^{+\infty}\frac{d\omega}{2\pi}\left[\frac{\partial}{\partial \omega}\ln G({\bf k},\omega)-G({\bf k},\omega)\frac{\partial}{\partial \omega}\Sigma({\bf k},\omega)\right]e^{i\omega t}
\end{eqnarray}
Here $\textrm{Tr}$ corresponds to trace over spin projections, and $\Sigma({\bf k},\omega)$ is the self energy. It would appear that the first term being a total derivative should vanish. However, one should be mindful that the time-ordered Green's function\footnote{Note that the retarded and the advanced Green's functions vanish at equal times.} is not analytic in either half plane with the chemical potential $\mu$ being the location where the poles switch from the upper-half to the lower-half plane. In the infinite volume limit, we have a cut along the entire real axis. For a Fermi liquid Luttinger argued that the second integral vanishes to all orders in perturbation theory, while the first integral leads to the relation 
\begin{equation}
N=\sum_{\bf k}\theta(\mu - \epsilon_{\bf k})
\end{equation}
where $\mu$ is the chemical potential for {\em interacting} electrons. How should one generalize this to the density wave states? This is trivial when one realizes that the trace operation must be enlarged to include the full Nambu space. The most general Hartree-Fock Hamiltonian that includes SDW, singlet DDW and the triplet DDW can be written in one stroke as
\begin{equation}
{\mathbb H}=\sum_{{\bf k}\in RBZ}{\mathbf \Psi}_{\bf k} ^{\dagger}{\mathbb A}_{\bf k}{\mathbf \Psi }_{\bf k} 
\end{equation}
where in the Nambu notation 
\begin{equation}
{\mathbf \Psi}_{\bf k} =\left( 
\begin{array}{c}
c_{\uparrow }\left( \mathbf{k}\right) \\ 
c_{\uparrow }\left( \mathbf{k+Q}\right) \\ 
c_{\downarrow }\left( \mathbf{k}\right) \\ 
c_{\downarrow }\left( \mathbf{k+Q}\right)
\end{array}
\right) .
\end{equation}
where ${\mathbb A}_{\bf k}$ is a $4\times 4$ matrix involving the respective order parameters. The corresponding Green's function can now be substituted in the Luttinger formula~\cite{Altshuler:1998}. The second term is obviously zero and one arrives at the result that 
\begin{equation}
x=x_{h}-x_{e}
\end{equation}
where $x$ stands for doped holes, while $x_{h}$ and $x_{e}$ stand for the concentration of carriers in the hole pockets and the electron pockets in the $RBZ$. Kohn's theorem applies once again to holes and electrons to the extent that interband transitions can be ignored. One would again begin with the unperturbed problem which fully includes the magnetic field resulting in Landau levels with gaps and turn on the residual interaction between the quasiparticles. The underlying assumption is that the Hartree-Fock gap is present, however small,\footnote{Magnetic breakdown effects may cause complication in the presence of large magnetic fields.}  and whatever residual interactions between the quasiparticles are present can be handled in exact analogy to the Luttinger's formalism. This result cannot persist across a quantum phase transition where the Hartree-Fock gap collapses.\footnote{Note that a topological Lifshitz transition is also a bonafide quantum phase transition.} A more complex question involves the behavior in the quantum critical region.

\section{Outstanding  puzzles}
\begin{itemize}
\item How does one reconcile with Fermi arcs observed in ARPES?
At the most trivial level one notes that once the coherence factors are incorporated for the ARPES spectral function~\cite{Chakravarty:2003}, as one must, the intensity of the parts of the Fermi surfaces become negligibly small, as shown in Fig.~\ref{fig:hole} and Fig.~\ref{fig:electron}.
\begin{figure}
\begin{center}
\begin{minipage}[htb]{5in}
\includegraphics[width=2 in]{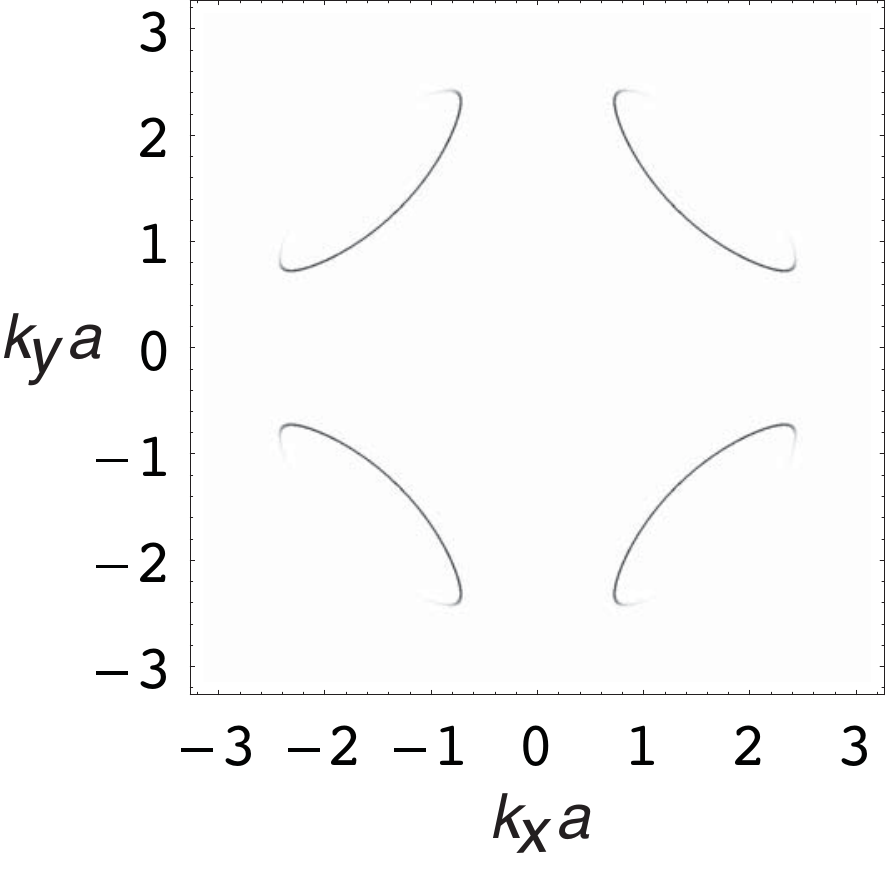} 
\hfill 
\includegraphics[width=2 in]{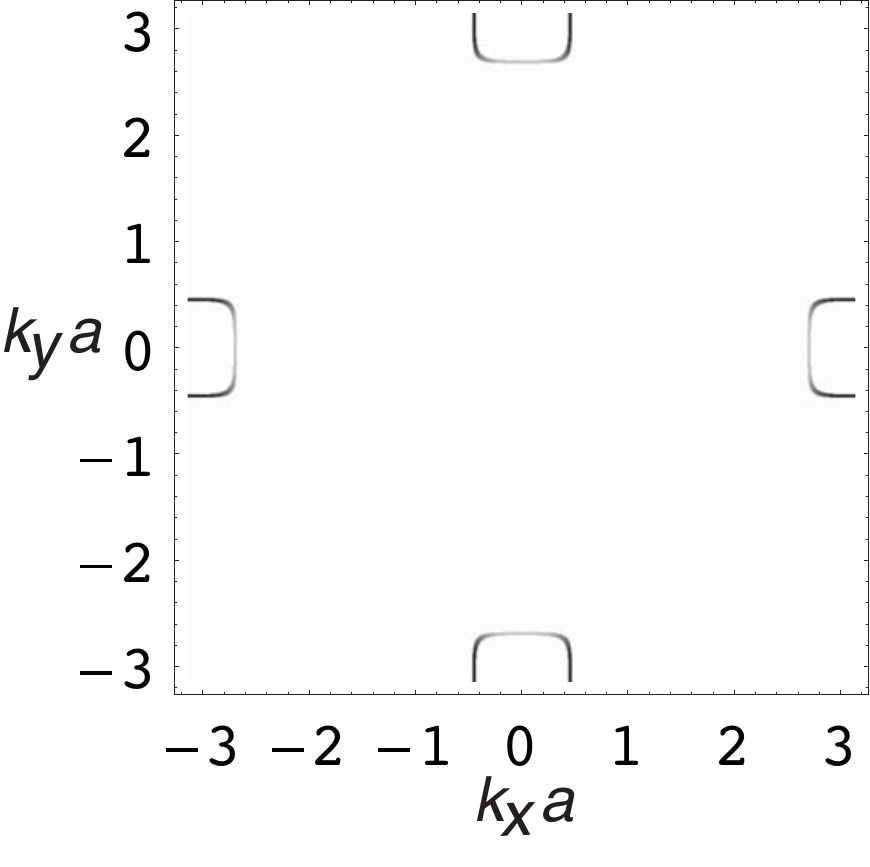}
\end{minipage}
\end{center}
\caption[hole]{\label{fig:hole}Left:  An example of calculated ARPES  spectral function without disorder for hole pockets where the effect of the coherence factors are prominent.}
\caption[electron]{\label{fig:electron}Right: An example of calculated ARPES spectral function without disorder for electron pockets.}
\end{figure}
At a deeper level, there are many unanswered questions. Although signatures of both hole and electron pockets in ARPES are known in electron doped superconductors~\cite{Armitage:2009}, there is no evidence so far of electron pockets in hole doped cuprates. But on the other hand there are no reliable ARPES measurements in YBCO. In a recent measurement~\cite{Hossain:2008}, it has been noted that a cleaved surface gets naturally overdoped regardless of what the bulk doping is. An attempt was made to reduce the doping by depositing a potassium overlayer and the ARPES showed no sign of electron pockets, nor even hole pockets, but only Fermi arcs. I have argued elsewhere that a likely picture is that potassium ions act to produce long-ranged disorder. The effect of this disorder is strikingly strong on electron pockets but less so on the hole pockets~\cite{Jia:2009}. Of course the effects of the coherence factors play some role as well. An example is shown in Figure~\ref{fig:disorder}.
\begin{figure}[htb]
\begin{center}
\includegraphics[scale=0.75]{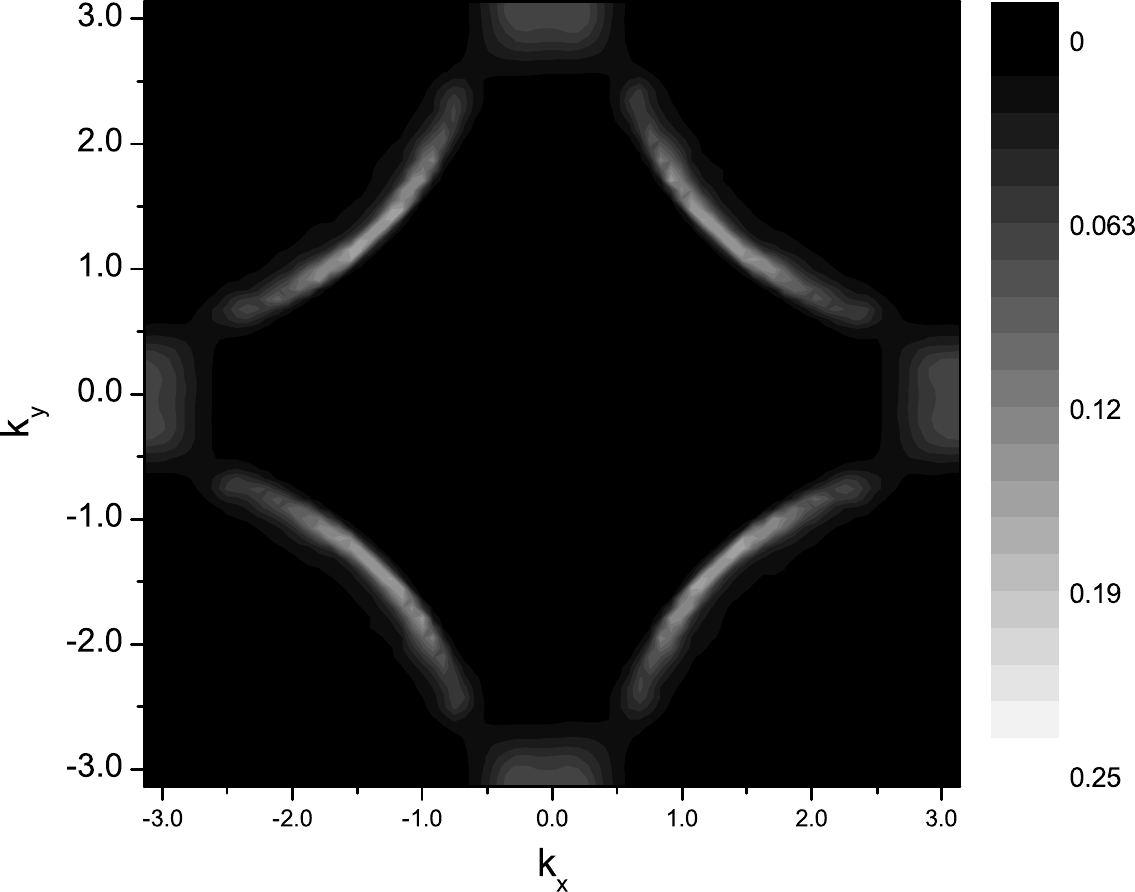}
\end{center}
\caption{An example of the effect of long ranged correlated disorder on the ARPES spectral function from Ref.~\cite{Jia:2009}. Note that the electron pockets are essentially wiped out and the visible part of the hole pocket is shortened.}
\label{fig:disorder}
\end{figure}
The definitive answer is missing, notwithstanding a recent ARPES measurement that finds both Fermi arcs and hole pockets~\cite{Meng:2009}.
\item Why is the hole pocket frequency that follows from Luttinger's sum rule is not observed in hole doped materials? At the most rudimentary level there is an answer. The Dingle factor, $D$,  that suppresses quantum oscillations can be written as 
\begin{equation}
D=e^{-p\pi/\omega_{c}\tau}
\end{equation}
where $p$ is the index  for the harmonic. Since $\omega_{c}= eB/m^{*}c$  and $v_{F}\tau=l$, where  $\tau$ the scattering time, and $l$ the mean free path,  $D$ can be alternately rewritten as 
\begin{equation}
D= e^{-p\pi\hbar c k_{F}/eBl}.
\end{equation}
Assuming that the mean free paths for the hole and the electron pockets are more or less the same, the pockets with larger $k_{F}$ will be strongly suppressed because of the exponential sensitivity. This appears to be reasonable from whatever numerical calculations and approximate analytical calculations exist~\cite{Jia:2009,Eun:2009}. Nonetheless, the definitive result in the resistive state of YBCO, including vortex fluctuations, is missing.
 \item Is the order that reconstructs the Fermi surface incommensurate? This perhaps will be settled experimentally. As to theory, presently we do not have sufficiently controlled microscopic understanding to address this question.
\item Do the tilted field measurements point truly to a triplet order parameter? This is also a matter that needs to be explored further experimentally. The implications, as emphasized earlier, are striking.
\item How about the existence of order and field induced order? At the moment there is no evidence of long-ranged order in the relevant doping range, be it SDW, singlet DDW, or triplet DDW, both in electron and hole doped cuprates. It is difficult to overemphasize that dHvA is an equilibrium measurement: fluctuating order cannot solve this dilemma, all quantum oscillation measurements require very large correlation lengths and nearly static  order. How about order induced by high magnetic fields?  There are strong arguments from detailed fits to the measurements that the relatively high magnetic field is not the root of these observations~\cite{Rourke:2009}, beyond the obvious effect of suppressing superconductivity. Indeed, previous NMR measurements in  $\mathrm{YBa_{2}Cu_{4}O_{8}}$ up to at least  $23.2\; \textrm{T}$ have shown no signatures of field induced order. Yet the quantum oscillation measurements in this stoichiometric material are clear and unambiguous. Of course, NMR measurements~\cite{Zheng:1999} in higher fields of the order of $45$T would be interesting. An argument that order could be field induced cannot be ruled out but seems to be implausible.

\item It is an open question at this time as to whether or not the quantum oscillation experiments can be understood using concepts of a non-Fermi liquid. These experiments have progressed to such a degree that there are many tight constraints imposed on theories.  If the tilted field experiment of Ref.~\cite{Ramshaw:2010} is sound, there appear to be unambiguous evidence of charge $e$, spin $1/2$, $m^{*}\sim 2$ fermions, consistent with a $g$-factor close to 2.2. I suspect that if an alternate explanation is possible, it will involve conceptual ideas similar to composite fermions at $\nu=1/2$, but with remarkable numerical coincidences.

\end{itemize}

\section{Outlook}

Challenging problems in the quantum theory of matter have been solved over the years  by startling  key ideas.
A fews specific examples are: (1) Landau theory of Fermi liquids.  Despite the fact that potential and kinetic energies are of similar magnitude, 
screening and the Pauli exclusion principle conspire
to allow a perturbative approach---the Landau Fermi liquid. Emergent quasiparticles  have renormalized mass, charge $e$, spin $1/2$ and weak residual interactions. (2) Critical phenomena---discovered by Thomas Andrews in 1869 was not explained until 1970's by Ken Wilson and his coworkers. Until then a nightmare of  fluctuations on all length scales confounded both experimentalists and theorists.  The key simplification was the notion of fixed points of renormalization group transformations. The physical systems that share the same fixed point have the same critical behavior. (3) BCS theory of superconductivity. With the key simplified reduced Hamiltonian  BCS were able to explain a myriad of mysteries and successfully predict new phenomena, far too many to recount here. The quagmire generated by high temperature superconductors {\em requires} simplification, not ill-defined complexity (akin to postmodernism), and a new concept of how a multiplicity of mechanisms can act in concert to raise the superconducting transition temperature in unconventional superconductors.

\ack
I thank Cyril Proust for much interesting correspondence. This work is supported by NSF under the Grant DMR-0705092.

\section*{References}

\begin{thebibliography}{10}

\bibitem{Doiron-Leyraud:2007}
Nicolas Doiron-Leyraud, Cyril Proust, David LeBoeuf, Julien Levallois,
  Jean-Baptiste Bonnemaison, Ruixing Liang, D.~A. Bonn, W.~N. Hardy, and Louis
  Taillefer.
\newblock Quantum oscillations and the {F}ermi surface in an underdoped
  high-$\mathrm{T_{c}}$ superconductor.
\newblock {\em Nature}, 447:565--568, 2007.

\bibitem{Jaudet:2008}
Cyril Jaudet, David Vignolles, Alain Audouard, Julien Levallois, D.~LeBoeuf,
  Nicolas Doiron-Leyraud, B.~Vignolle, M.~Nardone, A.~Zitouni, Ruixing Liang,
  D.~A. Bonn, W.~N. Hardy, Louis Taillefer, and Cyril Proust.
\newblock de {H}aas--van {A}lphen oscillations in the underdoped
  high-temperature superconductor $\mathrm{YBa_{2}Cu_{3}O_{6.5}}$.
\newblock {\em Phys. Rev. Lett.}, 100:187005--4, 2008.

\bibitem{LeBoeuf:2007}
David LeBoeuf, Nicolas Doiron-Leyraud, Julien Levallois, R.~Daou, J.~B.
  Bonnemaison, N.~E. Hussey, L.~Balicas, B.~J. Ramshaw, Ruixing Liang, D.~A.
  Bonn, W.~N. Hardy, S.~Adachi, Cyril Proust, and Louis Taillefer.
\newblock Electron pockets in the {F}ermi surface of hole-doped
  high-$\mathrm{T_{c}}$ superconductors.
\newblock {\em Nature}, 450:533--536, 2007.

\bibitem{Sebastian:2008}
Suchitra~E. Sebastian, N.~Harrison, E.~Palm, T.~P. Murphy, C.~H. Mielke,
  Ruixing Liang, D.~A. Bonn, W.~N. Hardy, and G.~G. Lonzarich.
\newblock A multi-component {F}ermi surface in the vortex state of an
  underdoped high-$\mathrm{T_{c}}$ superconductor.
\newblock {\em Nature}, 454:200--203, 2008.

\bibitem{Yelland:2008}
E.~A. Yelland, J.~Singleton, C.~H. Mielke, N.~Harrison, F.~F. Balakirev,
  B.~Dabrowski, and J.~R. Cooper.
\newblock Quantum oscillations in the underdoped cuprate
  $\mathrm{YBa_{2}Cu_{4}O_{8}}$.
\newblock {\em Phys. Rev. Lett.}, 100:047003--4, 2008.

\bibitem{Sebastian:2009}
Suchitra~E. Sebastian, N~Harrison, C.~H. Mielke, Ruixing Liang, D.~A. Bonn,
  W.~N. Hardy, and G.~G Lonzarich.
\newblock Spin-order driven {F}ermi surface revealed by quantum oscillations in
  an underdoped high-$\mathrm{T_{c}}$ superconductor.
\newblock {\em Phys. Rev. Lett.}, 103:256405, 2009.

\bibitem{Audouard:2009}
Alain Audouard, Cyril Jaudet, David Vignolles, Ruixing Liang, D.A. Bonn, W.N.
  Hardy, Louis Taillefer, and Cyril Proust.
\newblock Multiple quantum oscillations in the de {H}aas--van {A}lphen spectra
  of the underdoped high-temperature superconductor
  $\mathrm{YBa_2Cu_3O_{6.5}}$.
\newblock {\em Phys. Rev. Lett.}, 103:157003, 2009.

\bibitem{Vignolle:2008}
B.~Vignolle, A.~Carrington, R.~A. Cooper, M.~M.~J. French, A.~P. Mackenzie,
  C.~Jaudet, D.~Vignolles, Cyril Proust, and N.~E. Hussey.
\newblock Quantum oscillations in an overdoped high-$\mathrm{T_c}$
  superconductor.
\newblock {\em Nature}, 455:952--955, 2008.

\bibitem{Bangura:2008}
A.~F. Bangura, J.~D. Fletcher, A.~Carrington, J.~Levallois, M.~Nardone,
  B.~Vignolle, P.~J. Heard, N.~Doiron-Leyraud, D.~LeBoeuf, L.~Taillefer,
  S.~Adachi, C.~Proust, and N.~E. Hussey.
\newblock Small {F}ermi surface pockets in underdoped high temperature
  superconductors: Observation of {S}hubnikov--de {H}aas oscillations in
  $\mathrm{YBa_{2}Cu_{4}O_{8}}$.
\newblock {\em Phys. Rev. Lett.}, 100:047004--4, 2008.

\bibitem{Singleton:2009}
J.~Singleton, C.~De~La~Cruz, R.~D. McDonald, S.~Li, M.~Altarawneh, P.~Goddard,
  I.~Franke, D.~Rickel, C.~H. Mielke, X.~Yao, and P.~Dai.
\newblock Magnetic quantum oscillations in $\mathrm{YBa_2Cu_3O_{6.61}}$ and
  $\mathrm{YBa_2Cu_3O_{6.69}}$ in fields of up to $\mathrm{85 T}$; patching the
  hole in the roof of the superconducting dome.
\newblock {\em Phys. Rev. Lett.}, page 086403, 2010.

\bibitem{Rourke:2009}
P.~M.~C. Rourke, A.~F. Bangura, C.~Proust, J.~Levallois, N.~Doiron-Leyraud,
  D.~LeBoeuf, L.~Taillefer, S.~Adachi, M.~L. Sutherland, and N.~E. Hussey.
\newblock Fermi surface reconstruction and two-carrier modeling of the {H}all
  effect in $\mathrm{YBa_2Cu_4O8}$.
\newblock {\em http://arxiv.org/abs/0912.0175}, 2009.

\bibitem{Helm:2009}
T.~Helm, M.~V. Kartsovnik, M.~Bartkowiak, N.~Bittner, M.~Lambacher, A.~Erb,
  J.~Wosnitza, and R.~Gross.
\newblock Evolution of the {F}ermi surface of the electron-doped
  high-temperature superconductor $\mathrm{Nd_{2-x}Ce_{x}CuO_{4}}$ revealed by
  {S}hubnikov--de {H}aas oscillations.
\newblock {\em Phys. Rev. Lett.}, 103:157002--4, 2009.

\bibitem{Chakravarty:2008}
S.~Chakravarty.
\newblock Physics - {F}rom complexity to simplicity.
\newblock {\em Science}, 319:735--736, 2008.

\bibitem{Anderson:1997}
P.~W. Anderson.
\newblock {\em The theory of superconductivity in the high-$\mathrm{T_c}$
  cuprates}.
\newblock Princeton University Press, Princeton, New Jersey, 1997.

\bibitem{Kopp:2007}
Angela Kopp, Amit Ghosal, and Sudip Chakravarty.
\newblock Competing ferromagnetism in high-temperature copper oxide
  superconductors.
\newblock {\em Proc. Natl. Acad. Sci. USA}, 104:6123--6127, 2007.

\bibitem{Damascelli:2003}
Andrea Damascelli, Zahid Hussain, and Zhi-Xun Shen.
\newblock Angle-resolved photoemission studies of the cuprate superconductors.
\newblock {\em Rev. Mod. Phys.}, 75:473, 2003.

\bibitem{Norman:1998}
M.~R. Norman, H.~Ding, M.~Randeria, J.~C. Campuzano, T.~Yokoya, T.~Takeuchi,
  T.~Takahashi, T.~Mochiku, K.~Kadowaki, P.~Guptasarma, and D.~G. Hinks.
\newblock Destruction of the {F}ermi surface underdoped high-$\mathrm{T_c}$
  superconductors.
\newblock {\em Nature}, 392:157--160, 1998.

\bibitem{Gunnarsson:1997}
O.~Gunnarsson.
\newblock Superconductivity in fullerides.
\newblock {\em Rev. Mod. Phys.}, 69:575, 1997.

\bibitem{Chakravarty:1991}
Sudip Chakravarty, Martin~P. Gelfand, and Steven Kivelson.
\newblock Electronic correlation effects and superconductivity in doped
  fullerenes.
\newblock {\em Science}, 254:970--974, 1991.

\bibitem{Takabayashi:2009}
Yasuhiro Takabayashi, Alexey~Y. Ganin, Peter Jeglic, Denis Arcon, Takumi
  Takano, Yoshihiro Iwasa, Yasuo Ohishi, Masaki Takata, Nao Takeshita, Kosmas
  Prassides, and Matthew~J. Rosseinsky.
\newblock The disorder-free non-{BCS} superconductor $cs_3c_{60}$ emerges from
  an antiferromagnetic insulator parent state.
\newblock {\em Science}, 323:1585--1590, 2009.

\bibitem{Nguyen:2002}
Hoang~K. Nguyen and Sudip Chakravarty.
\newblock Effects of magnetic field on the $d$-density-wave order in the
  cuprates.
\newblock {\em Phys. Rev. B}, 65:180519, 2002.

\bibitem{Eun:2009}
Jonghyoun Eun, Xun Jia, and Sudip Chakravarty.
\newblock The surprising quantum oscillations in electron doped high
  temperature superconductors.
\newblock {\em arXiv:0912.0728}, 2009.

\bibitem{Li:2007}
Lu~Li, Yayu Wang, M.~J. Naughton, Seiki Komiya, Shimpei Ono, Yoichi Ando, and
  N.~P. Ong.
\newblock Magnetization, nernst effect and vorticity in the cuprates.
\newblock {\em Journal of Magnetism and Magnetic Materials}, 310:460--466,
  2007.

\bibitem{Chakravarty:2008b}
Sudip Chakravarty and Hae-Young Kee.
\newblock Fermi pockets and quantum oscillations of the {H}all coefficient in
  high-temperature superconductors.
\newblock {\em Proc. Natl. Acad. Sci. USA}, 105:8835--8839, 2008.

\bibitem{Millis:2007}
Andrew~J. Millis and M.~R. Norman.
\newblock Antiphase stripe order as the origin of electron pockets observed in
  1/8-hole-doped cuprates.
\newblock {\em Phys. Rev. B}, 76:220503--4, 2007.

\bibitem{Ramshaw:2010}
B.~J. Ramshaw, B.~Vignolle, R~Liang, W.~N. Hardy, C.~Proust, and D.~A. Bonn.
\newblock Angle-dependence of quantum oscillations in
  $\mathrm{YBa_2Cu_3O_{6.59}}$ shows free spin behaviour of quasiparticles.
\newblock {\em arXiv:1004.0260}, 2010.

\bibitem{Garcia:2010}
David Garcia-Aldea and Sudip Chakravarty.
\newblock Unpublished, 2010.

\bibitem{Kivelson:2003}
S.~A. Kivelson, I.~P. Bindloss, E.~Fradkin, V.~Oganesyan, J.~M. Tranquada,
  A.~Kapitulnik, and C.~Howald.
\newblock How to detect fluctuating stripes in the high-temperature
  superconductors.
\newblock {\em Rev. Mod. Phys.}, 75:1201--1241, 2003.

\bibitem{Sachdev:2003}
S.~Sachdev.
\newblock Colloquium: Order and quantum phase transitions in the cuprate
  superconductors.
\newblock {\em Rev. Mod. Phys.}, 75:913--932, 2003.

\bibitem{Nayak:2000}
C.~Nayak.
\newblock Density-wave states of nonzero angular momentum.
\newblock {\em Phys. Rev. B}, 62:4880--4889, 2000.

\bibitem{Chakravarty:2001}
S.~Chakravarty, R.~B. Laughlin, D.~K. Morr, and C.~Nayak.
\newblock Hidden order in the cuprates.
\newblock {\em Phys. Rev. B}, 63:094503, 2001.

\bibitem{Chakravarty:2002}
S.~Chakravarty.
\newblock Theory of the $d$-density wave from a vertex model and its
  implications.
\newblock {\em Physical Review B}, 66:224505, 2002.

\bibitem{Ghosal:2004}
Amit Ghosal and Hae-Young Kee.
\newblock Spatial variation of $d$-density wave order in the presence of
  impurities.
\newblock {\em Phys. Rev. B}, 69:224513, 2004.

\bibitem{Galitskii:1958}
V.~M. Galitskii and A.~B. Migdal.
\newblock Application of quantum field theory methods to the many body problem.
\newblock {\em Soviet Physics {JETP}}, 7:96--104, 1958.

\bibitem{Wilson:1983}
Kenneth~G. Wilson.
\newblock The renormalization group and critical phenomena.
\newblock {\em Rev. Mod. Phys.}, 55:583, 1983.

\bibitem{Yin:1996}
L.~Yin and S.~Chakravarty.
\newblock Spectral anomaly and high temperature superconductors.
\newblock {\em Int. J. Mod. Phys. B}, 10:805--845, 1996.

\bibitem{Mattis:1965}
D.~C. Mattis and E.~H. Lieb.
\newblock Exact solution of a many-fermion system and its associated boson
  field.
\newblock {\em J. Math. Phys.}, 6:304, 1965.

\bibitem{Nayak:1994a}
C.~Nayak and F.~Wilczek.
\newblock Renormalization-group approach to low-temperature properties of a
  non-fermi liquid-metal.
\newblock {\em Nucl. Phys. B}, 430:534--562, 1994.

\bibitem{Nayak:1994b}
C.~Nayak and F.~Wilczek.
\newblock Non-fermi liquid fixed-point in $(2+1)$ dimensions.
\newblock {\em Nucl. Phys. B}, 417:359--373, 1994.

\bibitem{Polchinski:1992}
J.~Polchinski.
\newblock Effective field theory and the fermi surface.
\newblock In J.~Harvey and J.~Polchinski, editors, {\em Recent Directions in
  Particle Theory}, Tasi, 1992. World Scinetific.

\bibitem{Shankar:1994}
R.~Shankar.
\newblock Renormalization-group approach to interacting fermions.
\newblock {\em Rev. Mod. Phys.}, 66:129, 1994.

\bibitem{Kohn:1961}
Walter Kohn.
\newblock Cyclotron resonance and de {H}aas-van {A}lphen oscillations of an
  interacting electron gas.
\newblock {\em Phys. Rev.}, 123:1242, 1961.

\bibitem{Luttinger:1960}
J.~M. Luttinger.
\newblock Fermi surface and some simple equilibrium properties of a system of
  interacting fermions.
\newblock {\em Phys. Rev.}, 119:1153, 1960.

\bibitem{Koulakov:1996}
A.~A. Koulakov, M.~M. Fogler, and B.~I. Shklovskii.
\newblock Charge density wave in two-dimensional electron liquid in weak
  magnetic field.
\newblock {\em Phys. Rev. Lett.}, 76:499, 1996.

\bibitem{Gell-Mann:1951}
Murray Gell-Mann and Francis Low.
\newblock Bound states in quantum field theory.
\newblock {\em Phys. Rev.}, 84:350, 1951.

\bibitem{Luttinger:1961}
J.~M. Luttinger.
\newblock Theory of the de {H}aas-van {A}lphen effect for a system of
  interacting fermions.
\newblock {\em Phys. Rev.}, 121:1251, 1961.

\bibitem{Abrikosov:1988}
A.~A. Abrikosov.
\newblock {\em Fundamentals of the theory of metals}.
\newblock Elsevier Science Publishers B. V., New York, 1988.

\bibitem{Wasserman:1991}
A.~Wasserman, M.~Springford, and F.~Han.
\newblock The de {H}aas-van {A}lphen effect in a marginal fermi-liquid.
\newblock {\em Journal of Physics-Condensed Matter}, 3:5335--5339, 1991.

\bibitem{Pelzer:1991}
Frank Pelzer.
\newblock Amplitude of the de {H}aas-van {A}lphen oscillations for a marginal
  {F}ermi liquid.
\newblock {\em Phys. Rev. B}, 44:293, 1991.

\bibitem{McCollam:2008}
A.~McCollam, J.~S. Xia, J.~Flouquet, D.~Aoki, and S.~R. Julian.
\newblock De {H}aas van-{A}lphen effect in heavy fermion compounds - effective
  mass and non-fermi-liquid behaviour.
\newblock {\em Physica B}, 403:717--720, 2008.

\bibitem{Ando:1974}
T.~Ando.
\newblock Theory of quantum transport in a 2-dimensional electron system under
  magnetic fields ({IV}). {O}scillatory conductivity.
\newblock {\em J. Phys. Soc. Jpn.}, 37:1233--1237, 1974.

\bibitem{Goswami:2008}
Pallab Goswami, Xun Jia, and Sudip Chakravarty.
\newblock Quantum oscillations in graphene in the presence of disorder and
  interactions.
\newblock {\em Phys. Rev. B}, 78:245406, 2008.

\bibitem{Leadley:1994}
D.~R. Leadley, R.~J. Nicholas, C.~T. Foxon, and J.~J. Harris.
\newblock Measurements of the effective mass and scattering times of composite
  fermions from magnetotransport analysis.
\newblock {\em Phys. Rev. Lett.}, 72:1906, 1994.

\bibitem{Du:1994}
R.~R. Du, H.~L. Stormer, D.~C. Tsui, L.~N. Pfeiffer, and K.~W. West.
\newblock Shubnikov-de {H}aas oscillations around $\nu = 1/2$ landau-level
  filling factor.
\newblock {\em Solid State Communications}, 90:71--75, 1994.

\bibitem{Jain:2007}
J.~K. Jain.
\newblock {\em Composite Fermions}.
\newblock Cambridge University Press, Cambridge, 2007.

\bibitem{Halperin:1993}
B.~I. Halperin, Patrick~A. Lee, and Nicholas Read.
\newblock Theory of the half-filled landau level.
\newblock {\em Phys. Rev. B}, 47:7312--7343, 1993.

\bibitem{Holstein:1973}
T.~Holstein, R.~E. Norton, and P.~Pincus.
\newblock de {H}aas-van {A}lphen effect and the specific heat of an electron
  gas.
\newblock {\em Phys. Rev. B}, 8:2649, 1973.

\bibitem{Wen:2004}
Xiao-Gang Wen.
\newblock {\em Quantum field theory of many-body systems}.
\newblock Oxford University Press, Oxford, 2004.

\bibitem{Altshuler:1994}
B.~L. Altshuler, L.~B. Ioffe, and A.~J. Millis.
\newblock Low-energy properties of fermions with singular interactions.
\newblock {\em Phys. Rev. B}, 50:14048, 1994.

\bibitem{Chakravarty:1995}
S.~Chakravarty, R.~E. Norton, and O.~F. Syljuasen.
\newblock Transverse gauge interactions and the vanquished fermi-liquid.
\newblock {\em Phys. Rev. Lett.}, 74:1423--1426, 1995.

\bibitem{Altshuler:1998}
B.~L. Altshuler, A.~V. Chubukov, A.~Dashevskii, A.~M. Finkel'stein, and D.~K.
  Morr.
\newblock Luttinger theorem for a spin-density-wave state.
\newblock {\em Europhys. Lett.}, 41:401--406, 1998.

\bibitem{Chakravarty:2003}
Sudip Chakravarty, Chetan Nayak, and Sumanta Tewari.
\newblock Angle-resolved photoemission spectra in the cuprates from the
  $d$-density wave theory.
\newblock {\em Phys. Rev. B}, 68:100504, 2003.

\bibitem{Armitage:2009}
N.~P. Armitage, P.~Fournier, and R.~L. Green.
\newblock Progress and perspectives on the electron-doped cuprates.
\newblock {\em arXiv:0906.2931}, 2009.

\bibitem{Hossain:2008}
M.~A. Hossain, J.~D.~F. Mottershead, D.~Fournier, A.~Bostwick, J.~L. McChesney,
  E.~Rotenberg, R.~Liang, W.~N. Hardy, G.~A. Sawatzky, I.~S. Elfimov, D.~A.
  Bonn, and A.~Damascelli.
\newblock In situ doping control of the surface of high-temperature
  superconductors.
\newblock {\em Nat. Phys.}, 4:527--531, 2008.

\bibitem{Jia:2009}
Xun Jia, Pallab Goswami, and Sudip Chakravarty.
\newblock Resolution of two apparent paradoxes concerning quantum oscillations
  in underdoped high-$\mathrm{T_{c}}$ superconductors.
\newblock {\em Phys. Rev. B}, 80:134503--8, 2009.

\bibitem{Meng:2009}
Jianqiao Meng, Guodong Liu, Wentao Zhang, Lin Zhao, Haiyun Liu, Xiaowen Jia,
  Daixiang Mu, Shanyu Liu, Xiaoli Dong, Jun Zhang, Wei Lu, Guiling Wang, Yong
  Zhou, Yong Zhu, Xiaoyang Wang, Zuyan Xu, Chuangtian Chen, and X.~J. Zhou.
\newblock Coexistence of {F}ermi arcs and {F}ermi pockets in a
  high-$\mathrm{T_c}$ copper oxide superconductor.
\newblock {\em Nature}, 462:335, 2009.

\bibitem{Zheng:1999}
Guo-qing Zheng, W.~G. Clark, Y.~Kitaoka, K.~Asayama, Y.~Kodama, P.~Kuhns, and
  W.~G. Moulton.
\newblock Responses of the pseudogap and d-wave superconductivity to high
  magnetic fields in the underdoped high-tc superconductor
  $\mathrm{YBa_{2}Cu_{4}O_{8}}$: An {NMR} study.
\newblock {\em Phys. Rev. B}, 60:R9947, 1999.

\end{thebibliography}

\end{document}